\documentclass[12pt,manuscript]{aastex}

\shorttitle{The SC-Type Star UY Cen}
\shortauthors{Steinfadt et al.}

\begin{document}

\title{A Unique Dust Formation Episode in the SC-Type Star UY Cen}
\author{Justin D. R. Steinfadt\altaffilmark{1,2}, Geoffrey C. Clayton\altaffilmark{2,3}, Tom Lloyd Evans\altaffilmark{4}, and Tom Williams\altaffilmark{5}}

\altaffiltext{1}{Department of Astronomy and Steward Observatory,
		 University of Arizona,
		 933 N Cherry Avenue, Rm. N204,
		 Tucson AZ 85721-0065; justins@physics.arizona.edu}
\altaffiltext{2}{Maria Mitchell Observatory,
                 Nantucket Maria Mitchell Association,
		 4 Vestal Street,
		 Nantucket MA 02554}
\altaffiltext{3}{Department of Physics and Astronomy,
		 Louisiana State University,
		 Baton Rouge LA 70803; gclayton@fenway.phys.lsu.edu}

\altaffiltext{4}{School of Physics and Astronomy,
		 University of St. Andrews,
		 St. Andrews, Fife, KY16 9SS, Scotland; thhle@st-andrews.ac.uk}
		 
\altaffiltext{5}{South African Astronomical Observatory, P.O. Box 9, Observatory 7935,
South Africa}
       
\begin{abstract}
We report the first detection of new dust formation in an SC star. The prototype of the SC stars, UY Cen, underwent a decline of 2 magnitudes in the V-band.  The SC stars show pulsational variations and have 60 \micron~excesses indicating past dust formation. It has been suggested that as a star evolves from oxygen rich to carbon rich, there is a short period of time when C/O$\sim$1 that the star appears spectroscopically as an SC star and ceases to produce dust. The SC star, BH Cru, has shown large spectroscopic and pulsation period variations in only 30 years, indicating rapid evolution but it has shown no sign of new dust formation. UY Cen has not shown any pulsation or spectroscopic variations accompanying the onset of its dust formation. In addition, UY Cen did not show emission in the resonance lines of Na I, K I or Rb I when it was at its faintest, although these lines were a feature of the carbon stars R Lep and V Hya during similar faint phases.

\end{abstract}
\keywords{stars: carbon --- stars: individual (\objectname{UY Cen}) --- stars: peculiar --- stars: variables: other}

\section{Introduction}
UY Cen is the "ur"-SC star. It was chosen as the prototype of its class when Catchpole \& Feast (1971) defined the SC stars. 
UY Cen was first discovered to be variable by Fleming who also noted that its spectrum was peculiar (Pickering 1911). Gaposchkin (1952) reported irregular photographic magnitude variations from 9.4 to 11.2 mag on Harvard College Observatory (HCO)  plates spanning the years 1904 to 1940. 

The SC stars are very rare. Only a few examples are known  (e.g., Catchpole \& Feast 1971; Keenan \& Boeshaar 1980). Even though they are very cool stars, they show almost no molecular bands but many atomic lines due to the fact that their C/O ratios are very near to 1. Spectroscopically, the SC stars can be characterized by very strong Na I D lines, a weak continuum to the blue of 4500 \AA, and the simultaneous presence of weak ZrO and CN bands (Catchpole \& Feast 1971; Keenan \& Boeshaar 1980; Wyckoff \& Clegg 1978; Za\u{c}s \& Sp\={e}lmanis 2001). 

SC stars typically have large 60 \micron~excesses indicating past episodes of mass-loss and dust formation (Jorissen \& Knapp 1998). However, their light variations are not due to dust obscuration.  They are all long-period pulsational variables with amplitudes ranging up to several magnitudes. Willems \& de Jong (1988) suggest that as an oxygen-rich AGB star begins to transition to its carbon-rich phase, and the C/O abundance ratio in the star approaches unity, mass loss greatly decreases and dust formation stops. Then, all the C and O atoms are locked up in CO molecules and are not available for dust formation (Jorissen \& Knapp 1998). The dust formation resumes when carbon becomes more abundant than oxygen. The transition timescale may be quite brief, between 10 and 10$^5$ yr (Willems \& De Jong 1988). 
One of the SC stars, BH Cru, has shown possible evolutionary changes. It
was classified as SC around 1970 (Keenan 1971; Catchpole \& Feast 1971) but
appeared as CS in 1980 (Lloyd Evans 1985a) and has consistently appeared
as CS or even a marginal C-type over the following two decades (Lloyd
Evans, to be published). Near infrared spectra in 1981 also suggested that
it was CS and not SC (Catchpole \& Whitelock 1985). Zijlstra et al (2004)
showed that the period has increased from 425 days in the early 1970s to
535 days recently. They suggest that a decline in effective
temperature, rather than a change in chemical composition, accounts for
the change from SC to CS.
The carbon Mira R Lep and the carbon semiregular V Hya both developed
strong emission in atomic resonance lines, as well as of the 5165 \AA~origin
band of the Swan system of C$_2$, during deep fadings, which were apparently
the result of dust formation and ejection, in the mid-1990s (Lloyd Evans
1997, 2000). 

Therefore, it was of great interest when UY Cen was recently found by Henshaw to be 2 magnitudes fainter than normal (Toone 2002). Its pulsation amplitude is only $\sim$0.4 mag. UY Cen had undergone a deep decline in visual brightness due to circumstellar dust formation. In this paper, we investigate observations of UY Cen over the past century to see whether it is just now evolving out of its SC phase.
We obtained new spectra of UY Cen in the relevant
spectral regions to compare with spectra taken in earlier years (e.g., Lloyd
Evans 1985b).

\section{Observations}

We examined the HCO archive for plates containing UY Cen. We found usable plates from four series, the A, MF, B, and RB. These are all blue plates. We found UY Cen on 139 plates from the period covering 1896 to 1953 with very irregular temporal coverage. The brightness of UY Cen was measured by eye through comparison with nearby stars in the field. The uncertainty in the m$_{pg}$ measurements is of the order of $\pm$0.2-0.3 mag, typical of such measurements (Schaefer 1994). 

Visual magnitudes of UY Cen were obtained from the AAVSO, Toone (2002), and from Albert Jones (personal communication). This dataset consists of 570  observations from 1989 to the present. 
UY Cen was also monitored by Hipparcos. The Tycho catalog contains 137 V-band observations between 1989 and 1993. 
The All Sky Automated Survey (ASAS) obtained 327 V-band observations of UY Cen from 2001-2005 (Pojmanski 2002).
There are a few other scattered observations in the literature (Eggen 1972; Walker 1979). All of these various observations were transformed to the V-band. 
For UY Cen, V$\sim$7.3 and B-V$\sim$2.8 (Eggen 1972).
For these values of V and B-V, the transformation from m$_{pg}$ to Johnson V [V-m$_{pg}$ = 0.17 - 1.09(B-V)] is quite large: V-m$_{pg}$ = -2.9 mag (Arp 1961; Pierce \& Jacoby 1995). The transformation from m$_{vis}$ to V has been determined to be m$_{vis}$-V = 0.21(B-V) (Stanton 1999). So for B-V = 2.8 mag, V-m$_{vis}$ = -0.6 mag. The photographic and visual magnitudes, transformed to V-band magnitudes, along with the V-band photometry are plotted in Figure 1. Individual Tycho photometric data taken on the same day were averaged. The AAVSO data were only used to fill the holes in the higher 
S/N ASAS data.
Gaposchkin (1952) gives 14 measures that are apparently averages of several HCO plates. He finds 
a variation of m$_{pg}$ = 9.4 to 11.2 (V= 6.5 to 8.3 mag) between 1904 and 1940. But, it is very unclear what plates were used so these measures aren't used in this study. 

To compare the brightness of UY Cen from different eras, we find that the average maximum light magnitude in the AAVSO data is m$_{vis}$=7.9 mag (V= 7.3 mag). This is very consistent with the magnitudes measured from the HCO plates. We find that the average brightness on the A, B and RB plates is m$_{pg}$ = 10.3, 10.1 and 10.0, respectively implying V$\sim$7.2 mag over the period from 1896 to 1953. On the MF plates, the average is m$_{pg}$ = 9.0 except in 1919 where it is m$_{pg}$ = 10.1 mag. There is enough overlap between the various plate series so that it is evident that the maximum light brightness of UY Cen is about 1 mag brighter on the MF plates than on the A, B and RB plates. 
Unfortunately, the color response on the MF plates is different from most of the other Harvard plates (Hoffleit 1979). Due to a sharp cutoff to the red of H$\beta$, if either the variable or comparison star is red, then there will be systematic errors. UY Cen is a very red star. 
Most of the HCO data are consistent with the measurement uncertainties but there are a few points (other than the MF plates) where UY Cen seems to be at V$\sim$6.5 mag. This is also seen in one photoelectric measurement (Eggen 1972). 

New JHKL photometry was obtained with the Mark II IR photometer on the 0.75m telescope at SAAO. These are listed in Table 1. Other JHKL photometry of UY Cen was found in the literature (Catchpole et al. 1979, 2MASS; Noguchi \& Akiba 1986; Fouqu\'{e} et al. 1992).  The K-L vs H-K and H-K vs J-H colors for the SAAO data are plotted in Figure 2. The 2MASS data are not plotted as their uncertainties are very large, $\sim$0.2-0.3 mag.

Figure 3 shows two visible spectra of UY Cen.  
The first spectrum was obtained on 1984 August 18 with the RPCS detector
and F1.4 camera with an 830 l/mm grating and slit width 300 microns (2\arcsec), giving a resolution of 
2 \AA~and S/N = 40, on the Cassegrain
spectrograph of the 1.9m reflector at SAAO, Sutherland. The second spectrum
was obtained on 2002 April 13 with the same spectrograph but using the CCD
detector and F2.2 camera with a 1200 l/mm grating and slit width 250
microns (1\farcs7) to give a resolution of 1.0 \AA~and S/N = 170. Spectra
of UY Cen were also obtained in red light, centered on 7600 \AA, with the same
spectrograph on 1999 June 2 and 2004 March 14. Both used the F2.2 camera,
1200 l/mm grating and slit width 250 \micron~to give resolution 1.4 \AA~and
S/N = 200, approximately.


\section{Discussion}

The pulsation period of BH Cru has been changing rapidly from 425 d in 1969 to 535 d in 1999 (Zijlstra et al. 2004). This period evolution is unique among the SC stars.
UY Cen shows no definitive period evolution in the data presented here. Gaposchkin (1952) suggested a period of 115 d based on fragmentary data from the HCO plates.  The small amplitude of the UY Cen pulsations, $\sim$0.4 mag, compared to BH Cru preclude using the HCO plate data for period determinations. However, the more modern data, shown in Figure 1, are fit well by a constant period of 176 d from 1989 to 2005. 
Koen \& Eyer (2002) find a period of 172.7 d and an amplitude of 0.21 mag from 95 Tycho (1989-1993) datapoints. However, when applied to more recent data from 2001 to 2005, 172.7 d is not a good fit. 
A period search on the 2001-2005 data was difficult due to the dust-formation brightness variations but gives a best fit to 181.8 d. This period does not fit the 1989-1993 data well. 
The value of 176 d fits well throughout both time periods but because of the large brightness variations due to dust formation as well as the pulsational variations, a better period cannot be ascertained at this time. 

Similarly, the spectrum of  BH Cru has been evolving rapidly (Zijlstra et al. 2004). The ZrO bands which were present before 1973 had disappeared by 1980 and were replaced by C$_2$ bands. No such spectral evolution seems to be taking place in UY Cen. 
In two high resolution spectra taken in 1993 and 1997, van Eck \& Jorissen (1999) note some small spectral changes in UY Cen.
Figure 3 shows the region of the D lines of Na I, as well as the 5635 \AA~
bandhead of C$_2$ and some weaker features of ZrO and YO. 
The 1984 spectrum of UY Cen shows weak ZrO at 6473 \AA~but no C$_2$ at 5635 \AA~(see Figures 1 and 2 of
Lloyd Evans 1985b); the 2002 spectrum again shows no C$_2$ at 5635 \AA, but the ZrO bandhead was not covered. However, several characteristic features of S stars which Catchpole (1980) noted as present on high-resolution spectra of UY Cen taken in 1954 and 1971, the weaker ZrO bandheads at 5627, 5718 and 5724 \AA~and the YO bandheads at 6132 and 6148 \AA, are present in both the
1984 and 2002 spectra. 
There is no significant difference between the absorption spectra of UY
Cen in 1984 and 2002 in this spectral region, given the factor two
improvement in resolution of the later spectrogram, which makes all weak
features appear stronger on average.
It is also apparent from Figure 3 that there was no
emission in the Na I D lines in 2002, when UY Cen was near its faintest point. The red spectra also bracket the fading, although the second spectrum was taken two years after the maximum fading,
by which time UY Cen was only 0.5 mag below normal brightness. The resonance lines of K I at 7699 \AA~and of Rb I at 7800 \AA~were measured on both spectra, yielding EW of 1.53 \AA~and 1.44 \AA~for
K I and 0.33 \AA~and 0.33 \AA~for Rb I at the earlier and later dates, respectively. We thus see no sign
of resonance fluorescence, nor of violet-displaced absorption, during or immediately following the
deep minimum of UY Cen. A spectrum of UY Cen is shown in Keenan \& McNeil (1976).

Willems \& de Jong (1988) have suggested that mass loss and dust formation stops in the SC-star phase when C/O approaches 1, i.e., all the C and O atoms are locked up in CO. When this happens,  the circumstellar dust shell moves away from the central star and cools as there is no new dust being added (Jorissen \& Knapp 1998). The star then makes a counterclockwise loop in the IRAS 25-60/12-25 color-color diagram. This color evolution ends when mass loss resumes as carbon becomes more abundant than oxygen. UY Cen and BH Cru may be on the lower part of the loop and about to enter the area of the color-color diagram where the IR carbon stars having large amounts of circumstellar dust are found (See Figure 13 in Jorissen \& Knapp 1998). 
The transition timescale for this loop may be quite brief, between 10 and 10$^5$ yr (Willems \& De Jong 1988). 
UY Cen shows horn-shaped CO profiles implying it has a detached shell and it also shows a 60 \micron~excess, indicating it has been experiencing a period of no mass loss
(Sahai \& Liechti 1995; Jorissen \& Knapp 1998).

As seen in Figure 1, the dust formation around UY Cen began sometime between 1999 March and 2000 December. The first photometry seems to have caught UY Cen at the end of a slow decline to a minimum which occurred around the beginning of 2002 when the star was about 2 mag below maximum. The brightness of UY Cen then  increased by about 1.5 mag levelling off in 2003 at about 0.5 mag below maximum light.  It remains about half a magnitude below its normal maximum light today.  JHKL photometry shows a similar evolution due to new dust around UY Cen. 
The K-L vs H-K and H-K vs J-H color-color diagrams are plotted in Figure 3. The colors of UY Cen from photometry obtained between 1975 and 1977 show very little variation.
The IR colors of UY Cen in 2002 during the major decline are consistent with an additional extinction of A$_V$ $\sim$2.5-3.0 mag. This is close to the amount of extinction seen at V, considering that the extinction was assumed to follow the typical ISM wavelength dependence  (Cardelli, Clayton \& Mathis 1989).
The most important difference between UY Cen and the other SC-type stars is that no other SC-type has undergone a documented decline like the one of UY Cen. The other SC-type stars, including BH Cru, are very regular periodic variable stars.
Feast et al. (1982) made a comparative study of the JHKL colours of a wide
range of red variable stars. The semiregular SC stars, including UY Cen,
all fall in a very narrow range of colors; the colors we observed for UY
Cen in 2002 March fall far to the red. The 2MASS database contains
single-epoch JHK photometry for a number of SC stars, which have similar
colors to UY Cen before the fading.

We searched for previous declines due to dust formation in the HCO plate data from 1886-1953. There is only one epoch in our plate data where we find a possible decline. In 1919, on the MF plates, UY Cen appears to be $\sim$1.5 mag fainter than on other plates of the same series.  The decline in 1919 may be a real dust formation event since we see a variation of 1.5 mag comparing UY Cen on just the MF Series plates. Unfortunately, since the MF plates are problematic and we can't properly place them in the context of the other measurements, this possible second decline due to dust remains uncertain.

Two carbon-rich long period variables, R Lep (Mira, P = 427 d) and V Hya (SR, P = 530 d, with
regular fadings at intervals of 18.3 years) were observed spectroscopically through dust-induced
light minima (Lloyd Evans 1997, 2000). Emission of C$_2$ bands and the resonance lines of Na I,
K I and Rb I were observed at and after minimum, while violet displaced absorption components of these lines appeared after minimum. Comparison of the spectra of UY Cen in 1984 and 2002
(Figure 3) shows no sign of emission in the Na I D lines even at the deepest stage of the minimum in 2002. The spectrum of March 2004, taken when UY Cen was only 0.5 mag fainter than normal, shows neither emission nor violet-displaced absorption of K I and Rb I. The significance of these observations is uncertain, as the minimum was not quite so deep as those of R Lep and especially
V Hya, and a collection of spectra of fluorescent emission of the resonance lines in cool stars of various types obtained by TLE shows that the ratio of the emission line strengths in the various lines may vary greatly between one star and another.

The detection of the resumption of mass loss and dust formation in UY Cen is an exciting development which may presage its evolution away from the SC-star stage. It is also possible that the suggestion of 
 Willems \& De Jong (1988) that mass loss stops during the SC-star phase is incorrect. Future monitoring of the spectrum and brightness of UY Cen may resolve this question. 

\acknowledgements
The authors would like to thank Dr. Vladimir Strelnitski, director of the Nantucket Maria Mitchell Observatory, for making it possible for this research to be conducted as part of a Research Experience for Undergraduates program during the summer of 2003. This research has made use of the SIMBAD database, operated at CDS, Strasbourg, France and of the NASA/IPAC Infrared Science Archive, which is operated by the Jet Propulsion Laboratory, California Institute of Technology, under contract with the National Aeronautics and Space Administration. This project was supported by the NSF/REU grant AST-0097694 and the Nantucket Maria Mitchell Association and Observatory. We acknowledge with thanks the variable star observations from the AAVSO International Database contributed by observers worldwide and used in this research. Many thanks to Albert Jones for obtaining photometry of UY Cen. We would also like to thank Alison Doane for her help with HCO plate collection.
 We are grateful to June McCombie and Arfon Smith for providing data for this study. 
This paper uses observations made from
the South African Astronomical Observatory (SAAO).




\begin{figure*}
\figurenum{1}
\includegraphics[scale=1.0,angle=0]{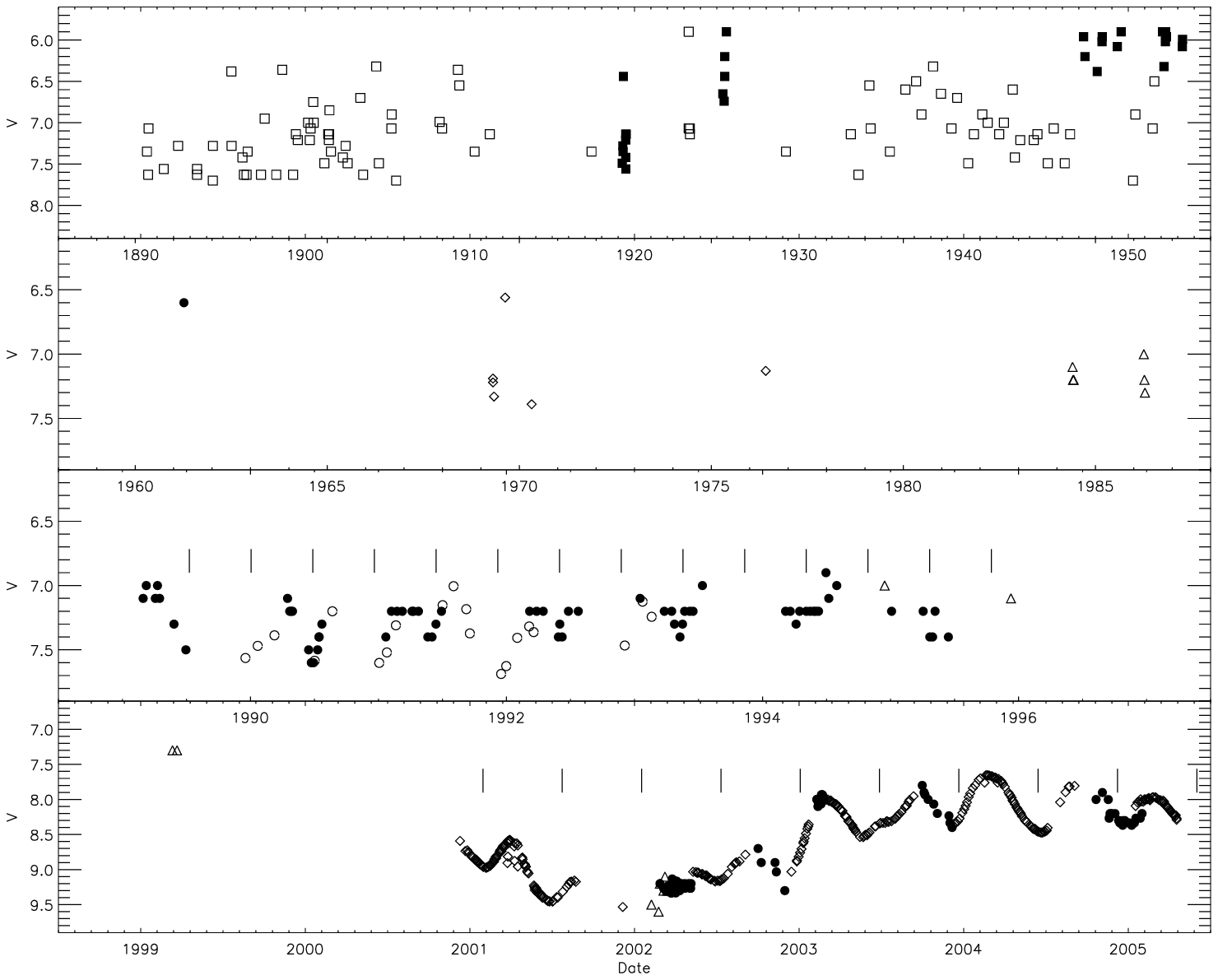}
\caption{Light curve for UY Cen from 1886 to 2005. Top panel: Magnitudes from HCO photographic plates. The MF plates are filled squares and the other plate series are plotted with open squares. In the remaining panels, the photometry plotted is, visual AAVSO data (filled circles), Tycho V-band photometry (open circles), other V-band photometry mostly from ASAS-3 (diamonds), and Toone's (2002) visual data (triangles). All of the data have been transformed to the V system. The time scales differ from panel to panel. 
The vertical bars indicate the 176 d pulsation period.}
\end{figure*}

\begin{figure*}
\figurenum{2}
\includegraphics[scale=1.0,angle=0]{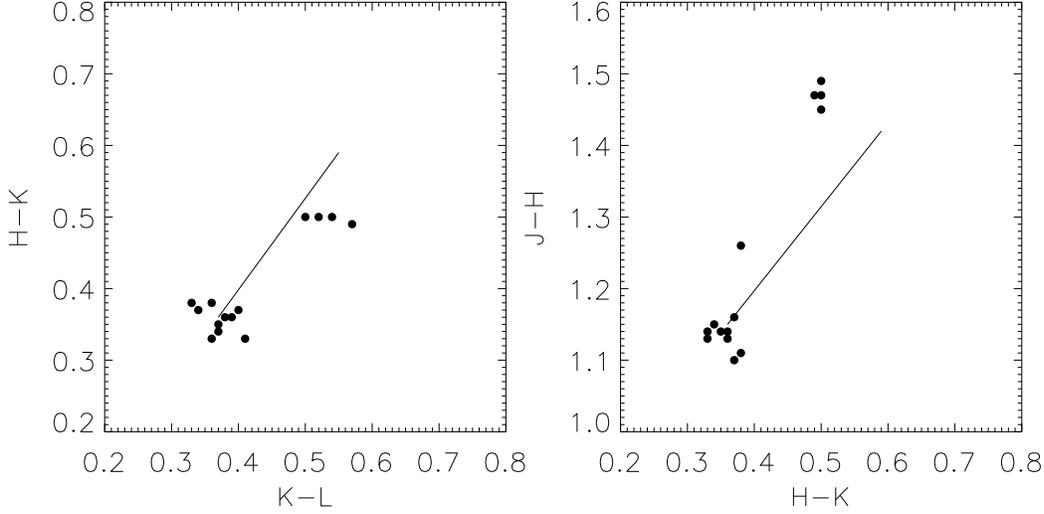}
\caption{IR color-color plots for UY Cen. Typical 1$\sigma$~uncertainties are 0.07 mag. In both diagrams, the points in the lower left are from 1975-1977, while those in the upper right are from 2002 after the episode of dust formation. The point midway between the two groups belongs temporally with the lower group and is not significantly different ($\sim2\sigma$). The lines represent CCM reddening of A$_V$=3 mag. }
\end{figure*}

\begin{figure*}
\figurenum{3}
\includegraphics[scale=0.7,angle=90]{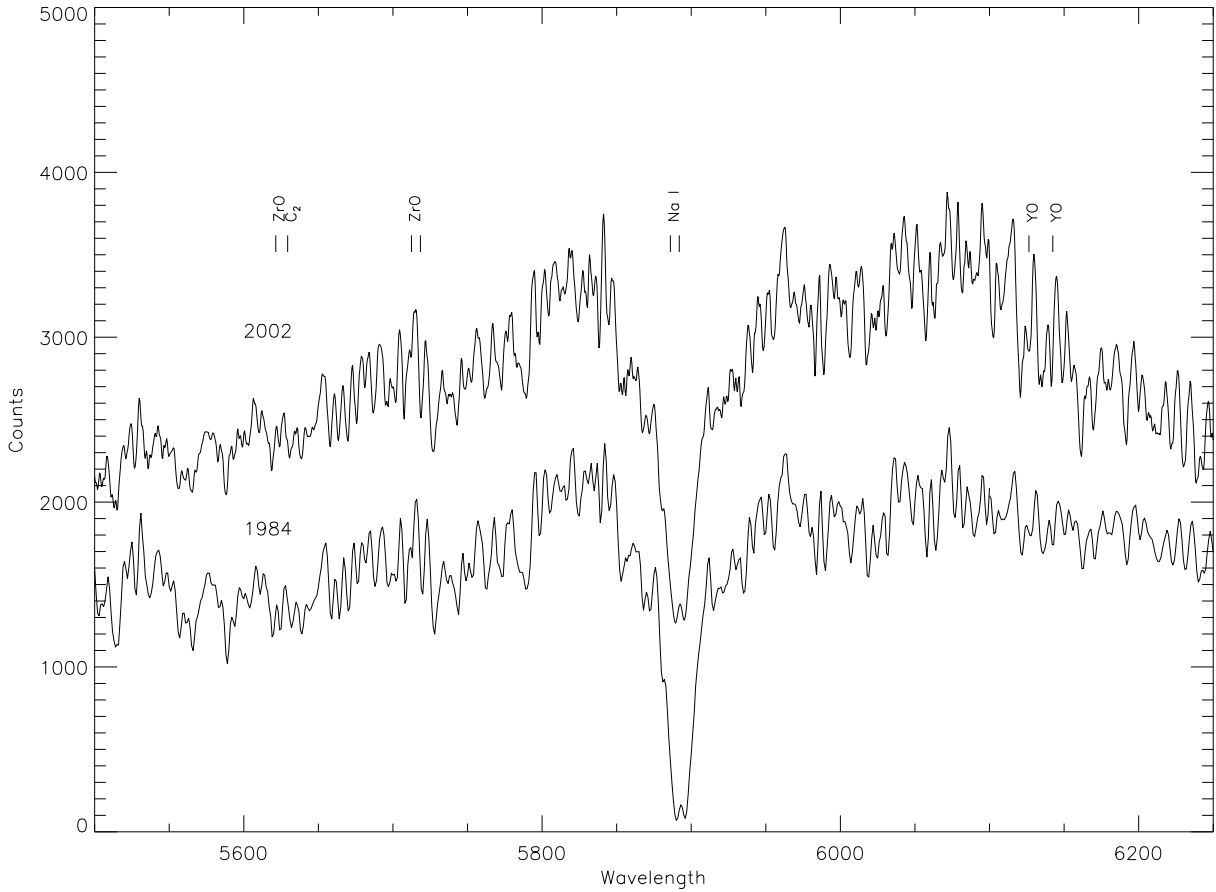}
\caption{Spectra of UY Cen taken in 1984 before the decline and in 2002 during the decline. There is
no significant difference between the spectra: in particular the weak ZrO and YO bands appear
unchanged. The Na I D lines are strong in absorption, as in all SC and CS stars, and there is no sign
of fluorescence emission within the Na I D lines in either spectrum.}
\end{figure*}

\begin{deluxetable}{lllll}




\tablenum{1}
\tablewidth{0pt}
\tablecaption{New JHKL Photometry of UY Cen}
\tablehead{\colhead{JD-245000} & \colhead{J} & \colhead{H} &
\colhead{K} & \colhead{L}}
\startdata
2360.50 &   2.99 $\pm$ .04  &   1.50  $\pm$ .04     &1.00  $\pm$ .04  &   0.50  $\pm$ .05\\
2363.50 &   2.95  $\pm$ .03   &  1.50  $\pm$ .03  &   1.00  $\pm$ .03    & 0.46  $\pm$ .04\\
2364.54  &  2.96  $\pm$ .03   &  1.49  $\pm$ .03  &   1.00  $\pm$ .03 &    0.43  $\pm$ .04\\
2365.53  &  2.95  $\pm$ .03   &  1.48  $\pm$ .03   &  0.98  $\pm$ .03  &   0.46  $\pm$ .04 ,
\enddata




\end{deluxetable}

\end{document}